\documentclass[aps,prl,showpacs,floats,twocolumn,floats,superscriptaddress,floatfix]{revtex4}

\usepackage{blindtext}

\usepackage{graphicx}
\usepackage{bm}
\usepackage{amsfonts}
\usepackage{color}

\usepackage{amsmath}    
\usepackage{epsfig}
\usepackage{subfigure}  
\usepackage{hyperref}   
\usepackage{bm}
\usepackage{amssymb}

\definecolor{gruen}{rgb}{0.1,0.5,0.1}
\definecolor{blau}{rgb}{0.1,0.2,0.8}

\newcommand{\be}{\begin{equation}}
\newcommand{\hr}{{\hat {\bf r}}}
\newcommand{\ee}{\end{equation}}

\newcommand{\hu}{\hat{u}}

\newcommand{\hbu}{\hat{{\boldsymbol u}}}
\newcommand{\hbf}{\hat{{\boldsymbol f}}}

\newcommand{\hbT}{\hat{{\boldsymbol T}}}

\def\bu{{\boldsymbol u}}

\newcommand{\bk}{{\boldsymbol k}}
\newcommand{\bx}{{\boldsymbol x}}

\newcommand{\br}{{\boldsymbol r}}

\newcommand{\romeaddress}{Department of Physics and INFN, University of 
Rome Tor Vergata,
Via della Ricerca Scientifica 1, 00133, Rome, Italy.}  
\newcommand{\nyuaddress}{{Department of Mechanical and Aerospace Engineering, New York University, New York, NY 11201, USA}}

\begin{document}

\title{Self-similar Subgrid-scale Models for Inertial Range Turbulence and Accurate Measurements of Intermittency
\footnote{postprint version of the manuscript published in Physical Review Letters 123.1: 014503 (2019).}}

\author{Luca Biferale}
\affiliation{\romeaddress}
\author{Fabio Bonaccorso} 
\affiliation{\romeaddress}
\author{Michele Buzzicotti} 
\affiliation{\romeaddress}
\author{Kartik P. Iyer} 
\affiliation{\nyuaddress}

\begin{abstract}
A class of spectral subgrid models based on a self-similar and reversible closure is studied with the aim to minimize the impact of subgrid scales on the inertial range of fully developed turbulence. In this manner, we improve the scale extension where anomalous exponents are measured by roughly one order of magnitude, when compared to direct numerical simulations or to other popular subgrid closures at
the same  resolution. {We found a first indication that intermittency for high order moments is not captured by many of the popular phenomenological models developed so far.} 
\end{abstract}
\maketitle

Turbulence is ubiquitous in nature and in engineering applications and it is characterized by the presence of intense non-Gaussian fluctuations on a wide range of inertial scales and frequencies.   The main mechanism  to be controlled and, eventually, modeled is the energy transfer from the large-scale, $L$,  where the flow is stirred, to the small-scale, $\eta$,  where viscous effects are dominant \cite{Frisch95,sreenivasan1999fluid,Pope00,ishihara2009study,alexakis2018cascades}. The Reynolds number is a measure of the separation between the two scales, $Re \sim (L/\eta)^{4/3}$. For most applications,  $Re$ is too large to allow the problem to be  attacked by direct numerical simulations (DNS) \cite{ishihara2009study,jimenez2012cascades}. Similarly, fundamental problems connected to the presence of anomalous scaling \cite{Frisch95,benzi2010inertial,iyer2017reynolds,sinhuber2017} 
in the limit  $Re \to \infty$ cannot be easily studied using numerical tools.
In such a deadlock, the applied community {resorts} to Large Eddy Simulations (LES), a numerical approach that restricts the Navier-Stokes equations to a range of scales (or wavenumbers) larger (smaller)  than  a given cut-off, $r > r_c$ ($k < k_c$), and modeling all subgrid-scale (SGS)  degrees of freedom with  closures in   configuration \cite{smagorinsky1963general,Pope00,meneveauARFM,lesieur2005large}, or Fourier  \cite{kraichnan1976eddy,chollet1981parameterization,baerenzung2008spectral,biferale2017optimal} space. The aim is to achieve a good accuracy for the energy-containing  modes, without paying too much attention to those (inertial) scales that are fully resolved, but also unavoidably affected by the subgrid-scale closure. As a matter of fact, most  LES implementation reproduce successfully the large-scale dynamics, $k \ll k_c$,  and are inaccurate for the highest resolved wavenumber modes, $k \sim k_c$.  This fact, prevents the possibility to use LES models to improve our understanding of multi-scale velocity fluctuations and/or the feedback of  small-scale fluctuations on global mean profiles. In particular,   SGS models (SGSM)  perform very poorly concerning the properties of the  inertial-range  scaling of velocity structure functions (SF):
\be
S_n(r)=\langle \left[ \delta_r u  \right]^n\rangle \sim \left( \frac{r}{L} \right)^{\zeta_n}
\label{eq:stfn_long}
\ee
where we defined the longitudinal increments $\delta_r u = [\bu(\br+\bx) -\bu(\bx)]\cdot \hr$  and we have assumed isotropy and homogeneity. The exponents $\zeta_n$ in  (\ref{eq:stfn_long}) are the key quantities to predict the asymptotic statistics for large Reynolds numbers, where $r/L$ can be arbitrarily small. 
\begin{figure}
\includegraphics[scale=0.55]{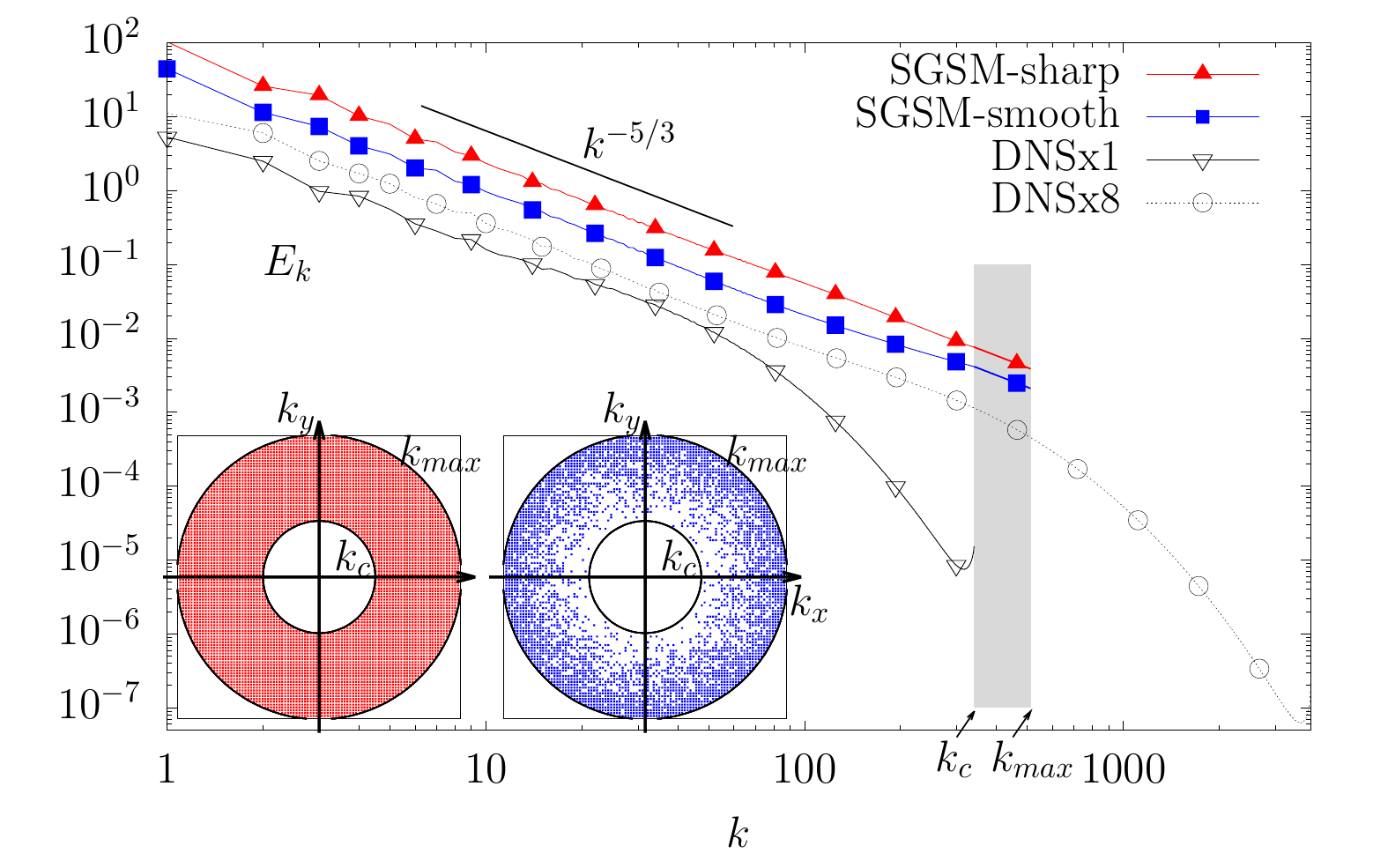}
\vspace{-0.6cm}
\caption{Energy spectra for the simulations in Table~\ref{tbl:table1}. The curves are shifted vertically for the sake of presentation. The grey area marks the range of wavenumbers where the closure acts. Inset: 2d sketch of the Fourier space support where $\gamma_\bk=1$. Left and right panels represent respectively SGSM-sharp and SGSM-smooth cases. 
}
\label{fig:1}
\end{figure}
On one side, experiments and numerical simulations have provided many evidences that the scaling of $S_n(r)$ is anomalous, i.e.  different from the Kolmogorov 1941 (K41) prediction $\zeta_n = n/3$ \cite{benzi2010inertial,gotoh2002velocity,watanabe2007inertial,sinhuber2017,iyer2017reynolds}.
On the other hand, we do not have any first-principle derivations of the  $\zeta_n$. Furthermore,  it is extremely difficult to get accurate measurements of the exponents, due to the concurrent requirements of having a large scaling range and large statistical ensembles. {As a result,  we also lack the  numerical and experimental accuracy to distinguish among different phenomenological models \cite{kolmogorov1962refinement,benzi1984multifractal,meneveau1987simple,she1994universal,schumacher2007asymptotic,eling2015anomalous,yakhot2001mean,yakhot2017emergence}. Finally, few assessments exist of the robustness of the exponents with respect to the small-scale dissipative mechanism \cite{falkovich1994bottleneck,lohse1995bottleneck,frisch2008hyperviscosity,donzis2010bottleneck}. 

In this letter, we introduce a class of subgrid models to minimize the impact of the SGS closure on the inertial-range: a sort of {perfect} energy-cascade sink  that achieves a much higher effective numerical resolution  to study  scaling  properties in turbulence. The idea was already presented in \cite{she1993constrained,jimenez1993energy} but was never applied and developed in the way it is here. We introduce a self-similar buffer close to the highest resolved mode, such as to have an ultraviolet boundary condition for the energy cascade at high $k$ which is consistent with the existence of an infinitely extended inertial range. The advantages with respect to other closures are many.  First, our model is time-reversible, allowing the formation of back-scatter events too. Second, it is a minor modification of the high-wave-number dynamics, without touching  the Fourier-phases and therefore with a minimal impact on the formation of intense coherent events that are believed to be the responsible of anomalous scaling. Unlike in \cite{jimenez1993energy}, here we focus on high Reynolds applications to assess the impact of the closure on the inertial range  properties. Furthermore, we expand the protocol by considering also a new Fourier modulation  where the closure is applied such as to improve its efficiency in absorbing the energy cascade.\\
In the following, we show that our LES protocol is able to obtain the same inertial-range extension of a fully resolved viscous DNS while saving roughly 1 order of magnitude of resolution. As a result, considering also the gain due to the possibility of relaxing the time step, the improvement in the computational resources is larger than a factor 1000, opening the way towards increased accuracy of measuring scaling exponents in turbulence, in both the scaling range extension and the statistical error.  Moreover, we assess the universality issue with respect to the ultraviolet dissipation mechanism by {comparing} the scaling obtained with our SGS-model with the ones measured in DNS and experiments \cite{gotoh2002velocity,sinhuber2017,iyer2017reynolds}. 
{Another by-product is to have a LES that is accurate for small-scale evolution, something  important
engineering applications that control extreme non-Gaussian events close to the subgrid
cutoff \cite{stevens2014large,stevens2017flow,buzzicotti2018effect,linkmann2018multi}.}
\\
\noindent
    {\sc The model.} Let us consider the Fourier-space evolution of the three dimensional Navier-Stokes equations in a periodic box of size $L=2\pi$ and resolved with $N$ grid points per direction and maximum wavenumber in all direction given by $k_{max}= N/2$:
\begin{equation}
\label{eq:NSE_F}
(\partial_t + \nu k^2) \hbu_{\bk}(t) = \hbT_{\bk}(t) + \hbf_{\bk}(t)
\end{equation}
where $\nu$ is the viscosity, $\hbf_\bk(t)$ is the Fourier transform of the external forcing and 
$\hbT_\bk(t) = - i\bk \cdot \left(\mathbb{I}-\frac{\bk \otimes\bk}{|k|^2} \right) \left[\sum_{\bk'} \hbu_{\bk'}(t) \otimes \hbu_{\bk'-\bk}(t) \right ]$  is the non-linear term.  We follow  \cite{jimenez1993energy} and we replace the viscous  term on the lhs of (\ref{eq:NSE_F}) with  a non-linear inertial closure that imposes a perfect self-similar Kolmogorov-like spectrum in a $k$-window close to the ultraviolet cut-off, $k_{max}$:
\begin{equation}
\label{eq:vincolo}
E_k(t) = \left(k/k_c\right )^{-\frac{5}{3}} E_{k_c}(t);\qquad k_c \le k \le k_{max},  
\end{equation}
where $E_k(t) =
\frac{1}{2}\sum_{|\bk|=k} |\hbu_\bk(t)|^2 $.
The LES equation for the resolved velocity field equipped with the fixed-spectrum SGS-model can be written using a Lagrangian multiplier
$\lambda_k(t)$ \cite{she1993constrained},
\be
\partial_t \hbu_{\bk}(t) = \hbT_{\bk}(t) + \hbf_{\bk}(t)- \gamma_\bk \lambda_k(t) \hbu_\bk(t)
\label{eq:LES_fixed}
\ee
where we have removed the viscosity and $\gamma_\bk$ is a projector which  selects the range of scales where the subgrid closure acts: $\gamma_\bk = 0 $ if $k \le k_c$ and  $\gamma_\bk = 1 $ if $k_c <k <k_{max}$ (SGSM-sharp).
It is easy to realize that in order to satisfy  (\ref{eq:vincolo}) we can impose $dE_k/dt =(k_c/k)^{5/3}d E_{k_c}/dt$ and choose   $\lambda_k(t)$ to be:
\be
\lambda_k(t) = \frac{1}{2} \frac{ T_k(t) - \left( k/k_c \right)^{-5/3} T_{k_c}(t)}{E_k(t)},
\label{eq:lagr_mult}
\ee
where $T_k(t)$ is the transfer function: $ T_k(t)= \sum_{|\bk|=k} \hbu_\bk^*(t) \hbT_\bk(t) $.  In order to mitigate the sharp transition  across the SGS, $k_c$,  we also explored another protocol where the percentage  of constrained modes grows linearly {from} $0$ at $k_c$ to $1$ at $k_{max}$. To do that,  we define a (quenched)  probability to apply the SGS model at any given wavenumber as follows (SGSM-smooth):
\be
\gamma_\bk = \begin{cases}
0 \text{ if } k < k_c  \nonumber\\
1 \text{ with prob.  } P_k = \frac{k-k_c}{k_{max}-k_c} \text{ if }  k_c \le k < k_{max}. \nonumber
\end{cases}
\ee
In this way, only a fraction of modes $(k-k_c)/(k_{max}-k_c)$ will be affected by the constraint for any given shell $k$, such that we move from fully unconstrained dynamics (for $k<k_c$) to  a fixed spectrum dynamics (for $k=k_c$) with continuity (see inset of Fig.~\ref{fig:1} for a graphical scheme of the Fourier space support of the projector $\gamma_\bk$ for both sharp and smooth SGSM cases). { We also anticipate that in order to minimize the transition across $k_c$ we will need to keep a small residual viscosity $\nu$ even when using the self-similar closure. This is unavoidable due to the fact that the closure acts on a finite range of scales and cannot mimic exactly the SGS dynamics at infinite Reynolds. }\\

\begin{table}
    \begin{tabular}{|c|ccc|c|c|c|c}
     & $N$ & $k_c$ & $k_{max}$  &  $\varepsilon$ & $\nu$ & $T$ & $Re$ \\
    \hline 
    SGSM-sharp  & $1024$ & 340 & 512   &  3.0 & $8.0 \cdot 10^{-5}$ & 8.5  & $2.1 \cdot 10^5$\\
SGSM-smooth & $1024$ & 340 & 512   &  3.0 & $4.0 \cdot  10^{-5}$ & 8.5  & $4.2 \cdot 10^5$ \\
    DNSx1      & $1024$ & ...  & 340   & 2.5 & $8.0 \cdot 10^{-4}$ & 12 & $2.0 \cdot 10^4$ \\
    DNSx8      & $8192$ & ...  & 3861 & 1.5 & $4.4 \cdot 10^{-5}$ & 3.4 & $3.0 \cdot 10^5$ \\
    \end{tabular}
    \caption{ Simulations:
    $N$: number of collocation points in each spatial direction; 
    $k_c$: smallest wavenumber where the SGS closure acts. 
    $k_{max}$: maximum wavenumber evolved by the dynamics.
    $\varepsilon$: mean energy injection;
    $\nu$: kinematic viscosity. $Re = \varepsilon^{1/3} L^{4/3}/\nu$: Reynolds number with $L=2\pi$. $T$: duration of simulations in units of the  eddy turn over time $\varepsilon^{-1/3}L^{2/3}$ }
    \label{tbl:table1}
\end{table}

{\sc Results}. We compare the LES data  obtained at a resolution of $1024^3$  with the two different DNS resolutions:  one identical to the LES (DNSx1) and one taken from a state-of-the-art study at  $8192^3$ collocation points \cite{iyer2017reynolds}  denoted (DNSx8). All runs are forced with a white-in-time Gaussian forcing acting at $k_f \in [1,1.5]$ for DNSx1 and a  $k_f \in [1,3]$  for DNSx8.
More details on the numerical set up can be found in Table I. 
In Fig.~\ref{fig:1} we show the spectral properties of all data. Our closure reproduces the same extension of the scaling range of DNSx8 and considerably extends the one obtained with DNSx1. 
{ We obtain an inertial
behavior for all $k$ in the LES model without the viscous range of scales needed 
with standard viscosity in DNSx8.} \\
{\sc Anomalous scaling of high order SF}. To assess the scaling properties in a quantitative way,  we  measure the local scaling exponents:
\be
\xi_n(r) = \frac{d \, \log S_n(r)}{d \, \log (r)}
\label{eq:local_slopes}
\ee
where in the presence of pure power-laws  we must have $\xi_n(r) = const. = \zeta_n$.
\begin{figure}[htp]
\includegraphics[scale=0.5]{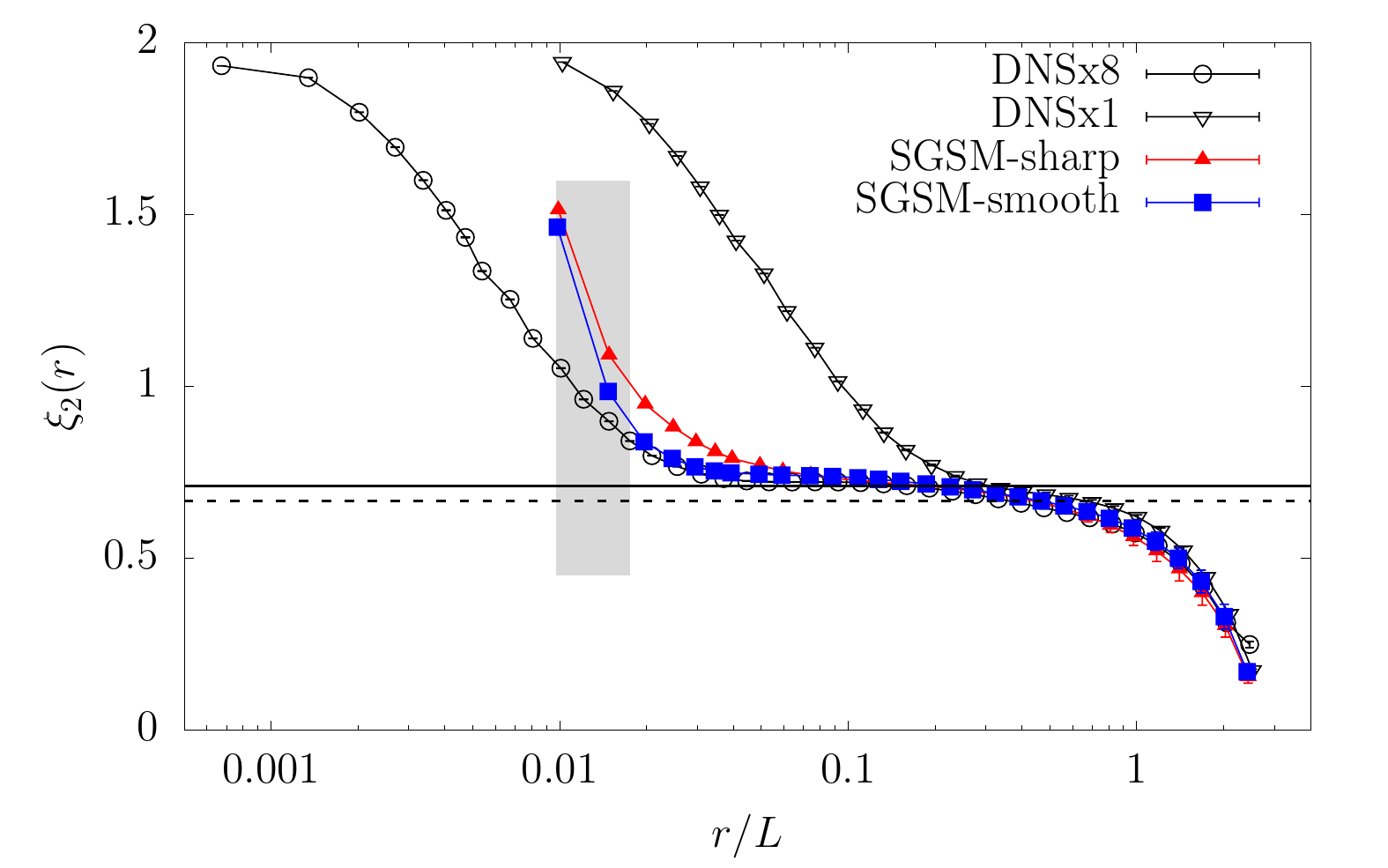}
\vspace{-0.3cm}
\caption{Log-lin plot of $\xi_2(r)$ vs $r$. Solid and dashed lines indicate the She-Leveque (SL) $\zeta_2=0.69$ and K41 $\zeta_2=2/3$ predictions, respectively. In grey, we indicate the range of scales where the closure (\ref{eq:lagr_mult}) is acting. Error bars are comparable with symbols' size. }
\label{fig:2}
\end{figure}
\begin{table*}
    \begin{tabular}{|c|c|c|c|c|c|c|}
      &  SGSM & DNSx1 & DNSx8 & SL  & Ya & EO\\
    \hline  
    {n=4}  & $1.843 (15)$ & $1.828 (25)$  & $1.824 (18)$ & $1.839$ & $1.843(15)$ & $1.843(15)$\\
    {n=6}  & $2.537 (38)$ & $2.501 (78)$  & $2.485 (39)$ & $2.555$ & $2.563(38)$ & $2.586(35)$\\
    {n=8}  & $3.092 (30)$ & $3.034 (147)$ & $2.982 (56)$ & $3.176$ & $3.186(66)$ & $3.257(58)$\\
    {n=10} & $3.504 (81)$ & $3.440 (230)$ & $ - $        & $3.727$ & $3.730(96)$ & $3.875(83)$\\
    \end{tabular}
    \caption{{$\bar{\Delta}_n+n/2$  obtained as a fit
  of $\xi_n(r)/\xi_2(r)$ for $r \in [0.03:0.9]\,L$ for SGSM-smooth and DNSx8, and for $r \in [0.15:0.9]\,L$ for DNSx1. Errors for the numerical data refer to the sum among  statistical fluctuations and the variations considered
      by fitting in the first or second half of the scaling range (see the Supplemental Material for more details). The last three columns give the prediction from  She-Leveque (SL) \cite{she1994universal}, Yakhot (Ya) \cite{yakhot2001mean} and Eling-Oz (EO) \cite{eling2015anomalous} models, where the last two have been fitted to have the value for $n=4$ identical to the smooth SGSM case. Errors in the Ya-EO models are estimated by fixing their free parameter to match either the maximum SGSM value $1.843+0.015$ or the minimum, $1.843-0.015$, for $n=4$, see Supplemental Material.}}
    \label{tbl:table2}
\end{table*}

By measuring where $\xi_n(r)$  is constant we have an unbiased definition of the inertial range extension and we can assess scale-by-scale the quality of our data. In particular, intermittency and scale-dependent corrections from a Gaussian behavior can be measured by the deviation from zero of  $\Delta_n(r) = \xi_n(r)/\xi_2(r)-n/2$ as  seen by expressing the generalized Flatness in terms of the 2nd order SF: {
$$
F_n(r) = \frac{S_n(r)}{[S_2(r)]^{n/2}} \sim [S_2(r)]^{\left( \frac{\xi_n(r)}{\xi_2(r)}-\frac{n}{2} \right)}.
$$}
In Fig.~\ref{fig:2} we show $\xi_2(r)$  for our two SGSM closures and compare them with the same quantity measured on DNSx1 and on DNSx8. 
\begin{figure}[htp]
\includegraphics[scale=0.55]{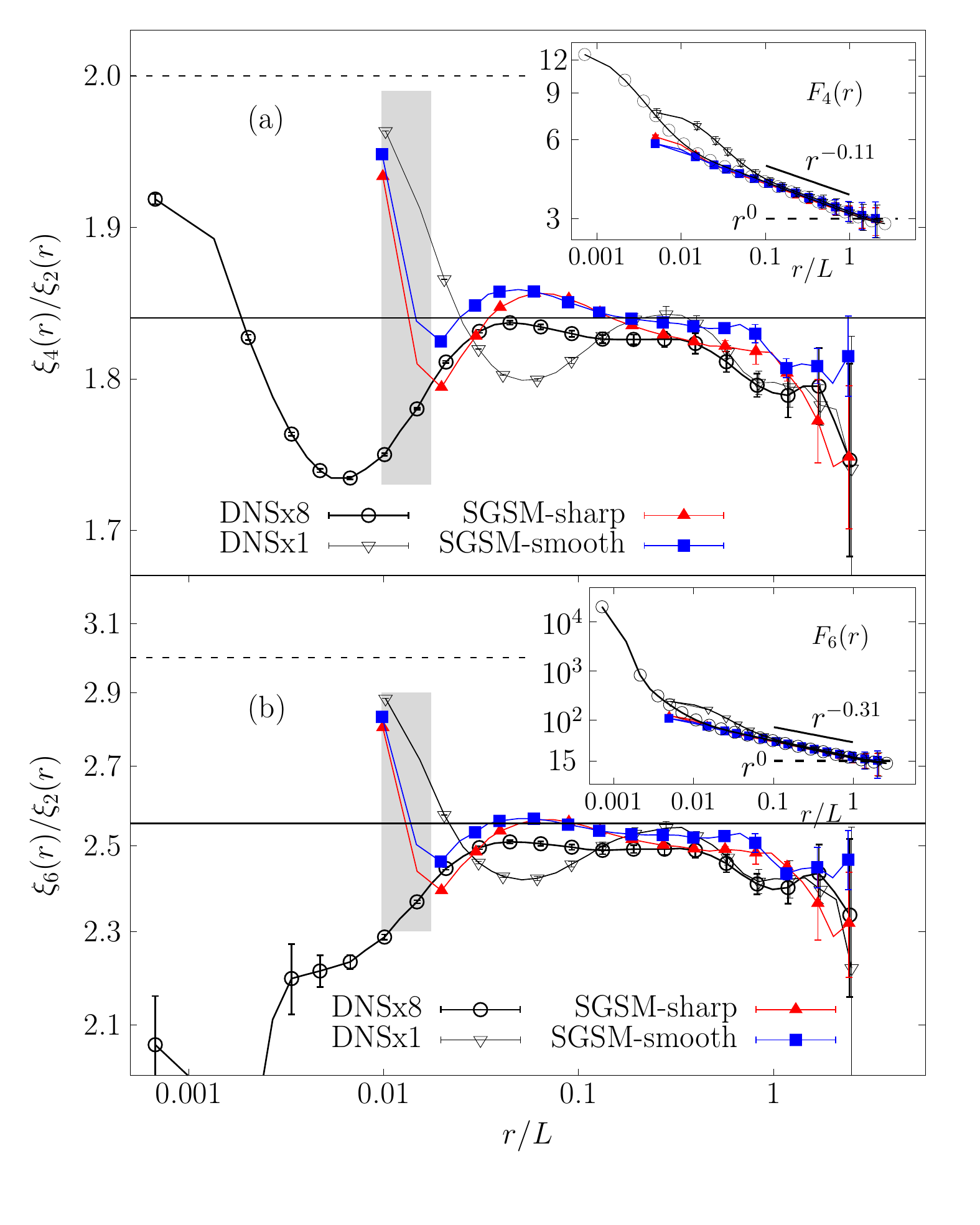}
\vspace{-1.cm}
\caption{Log-lin plot of $\xi_n(r)/\xi_2(r)$ for $n=4$ (top)  and $n=6$ (bottom) for  SGSM and DNS data. K41 and SL predictions are given by the dashed and solid lines respectively. EO and Ya models are very close to SL for these two moments (see table II).  In grey we indicate the range of scales where the closure (\ref{eq:lagr_mult}) is applied.
  Inset: log-log plot of $F_n(r)$ vs $r$ (same symbols of the main
  panel).  SL and K41 scaling are given by the solid and the dashed lines, respectively.   In all figures, errors are evaluated from the scatter of 40 configurations. }
\label{fig:3}
\end{figure}
As shown for the spectral case, LES data have a much larger extension of  scaling  then DNSx1, matching the DNS obtained with a 8-times larger  resolution (DNSx8). \\ {Despite the existence of a {\it plateau} for $\xi_2(r)$  for all data, the constraint  $\xi_2(r) \to 2$ for $r \to 0$ makes the jump from inertial to viscous values too big and it is very difficult to quantitatively distinguish the Kolmogorov 1941 (K41) scaling  from any  intermittent phenomenological model as, e.g. the She-Leveque (SL) \cite{she1994universal}, the Yakhot model \cite{yakhot2001mean} and the model proposed by Oz based on spontaneous symmetry breaking of dilation invariance and random geometry  \cite{oz2018turbulence,oz2018scale,eling2015anomalous}. {To be  more accurate, in Fig.~\ref{fig:3} we show the scaling of the generalized Flatness (inset) and of the scale-by-scale ratio $\Delta_n(r)+n/2= \xi_n(r)/\xi_2(r)$  (main panel) for $n=4,6$.} Here, a Kolmogorov-like nonanomalous scaling corresponds to a constant value $n/2$ for all $r$. As one can see, the deviation from the Kolmogorov scaling is now evident and -much more importantly- our SGSM closures are able to develop an inertial range as extended as the DNSx8 case, if not even larger. Moreover, the SGSM-smooth closure is a bit better than the SGSM-sharp case.  We consider these results  a clear demonstration that the SGS {model developed here} can be considered a sort of  infinite-Reynolds closure. {Considering the fact that using the SGSM-smooth closure we can achieve the same accuracy for local exponents of a DNS with  8-time larger resolution, we  estimate  a gaining factor $8^3$ for the spatial grid,  which together with the less stringent Courant-Friedrichs-Lewy (CFL) condition for the time integration, 
 $\sim \varepsilon^{-1/3}k_{max}^{-2/3}$ leads to a total  gain close to a factor 1000.} In Table II we  present a summary for the scaling properties of $F_n(r)$ { from where it is clear that the SGS models agrees with the DNSx8 and with the prediction made by models SL-Ya-EO for moment, $n=6,8$}   {while for the largest achievable order, $n=10$, numerical data are more intermittent than all three phenomenological models (see also SM).}

A few comments are now in order. { First, it is useful to preserve a very small viscous term in (\ref{eq:LES_fixed}) in order to have a  smooth transition across $k_c$. This is implemented in our approach,  keeping a term $\nu k^2 \hbu_{\bk}(t)$  with a very small $\nu$  as shown in Table I. It is clear from Fig.~\ref{fig:3} that even by optimizing $\nu$,  there exists in the SGSM a pseudo-{\it viscous}  range (extended over a few grid points) where scaling breaks down.  This is  unavoidable because our closure is acting in the Fourier space and does not enforce any pure scaling for the high order SFs. The existence of a small bump  for the local slopes around the transition from viscous to inertial range is present also in experimental data at high Reynolds \cite{sinhuber2017}.} On the other hand, the efficiency in extending the anomalous scaling-range  is a good evidence that to capture intermittency the SGSM must maintain the correct phase-correlations \cite{murray2018energy}, which is one of the main added value of (\ref{eq:lagr_mult}). {Second, the smooth projector recipe is not unique and one can imagine many different ways to enforce the transition from  modes that evolve according to their Euler dynamics ($k<k_c$) to those that feel the spectral constraint. In particular, once the controlled buffer is introduced and it is large enough, one might imagine even avoiding the dealiasing protocol and keeping $k_{max} = N/2$ as done here. The effects of introducing a dealiasing are minor and discussed in Fig. 2 of the SM.} 
\begin{figure}[htp]
\includegraphics[scale=0.55]{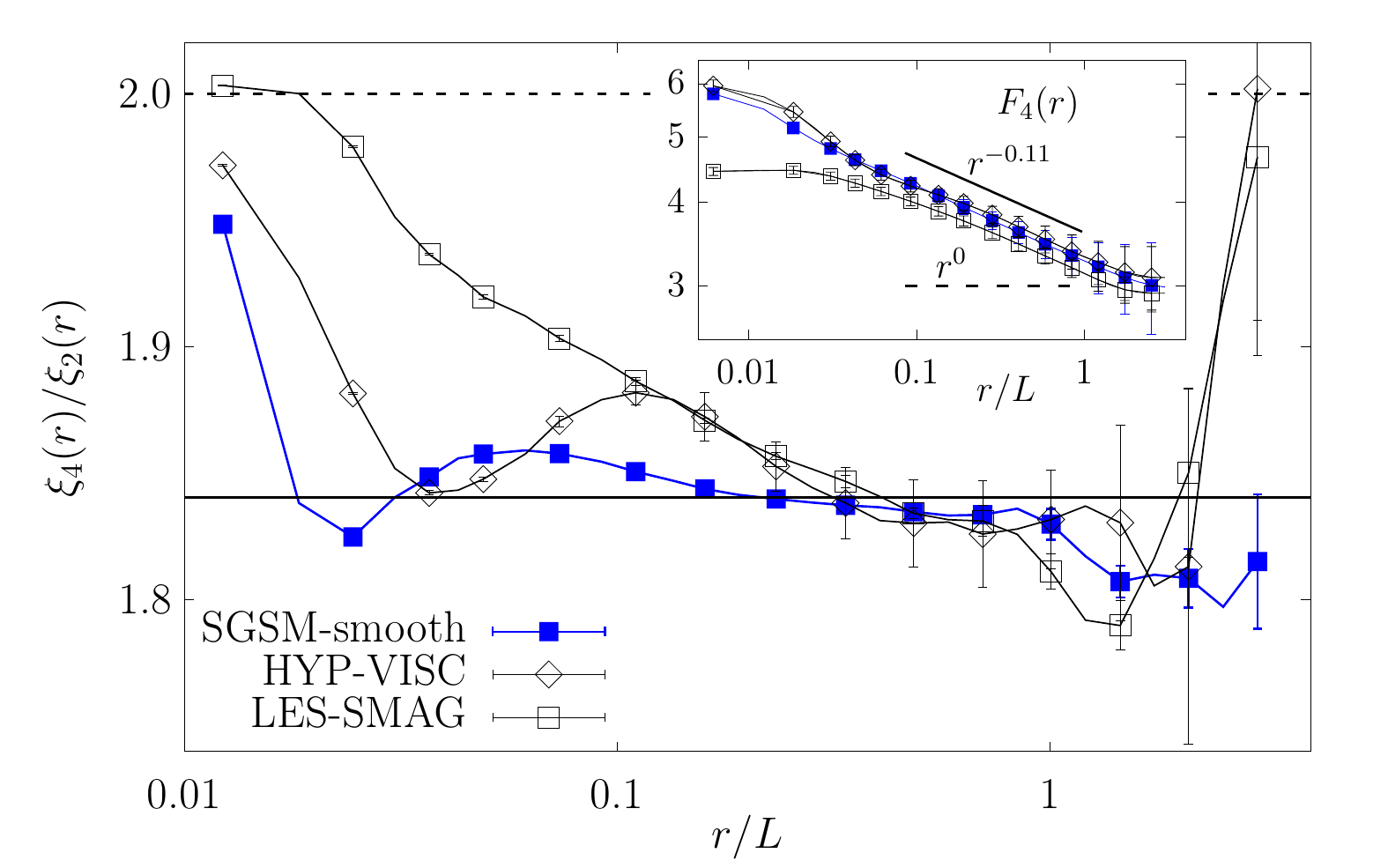}
\vspace{-0.6cm}
\caption{ Comparison of $\xi_4(r)/\xi_2(r)$ for (i) SGSM-smooth, (ii) Smagorinsky-LES and (iii)  hyper-viscous DNS with  $\nu\Delta^2 \bu$ and $\nu=2.0 \cdot 10^{-8}$. All simulations have $1024^3$ collocation points.  The SL and K41 prediction are given by the solid and the dashed lines, respectively.  Inset: $F_4(r)$ for the same data.}
\label{fig:4}
\end{figure}
We now discuss the comparison with two other popular ways to {\it enhance} the effective Reynolds numbers. In Fig.~\ref{fig:4} we compare the Flatness obtained from  a DNS with hyper-viscosity  \cite{borue1995self,frisch2008hyperviscosity} or from a  Smagorinsky SGS model \cite{smagorinsky1963general,meneveauARFM,linkmann2018multi} with the one proposed here. Notice that the hyperviscous data are only qualitatively as good as the SGSM-smooth  as shown by  the fact that the former has a less extended plateau wrt to the latter. There are no doubts that the closure \eqref{eq:lagr_mult} is superior to both  Smagorinsky and hyperviscous models. 
Finally,  we mention that from  (\ref{eq:LES_fixed}) one can define a Galilean-invariant \cite{eyink2009localness} SGS energy transfer:
$ \Pi(\bx) = \partial_i u_j (\bx)\int d\bk \,  \gamma_\bk \lambda_k  e^{i\bk\cdot \bx}
 ik_i/k^2 \hu_{j,\bk}  $ which is non-positive definite and therefore able to reproduce back-scatter events. 
\\ 
{\sc Conclusions}. We have shown that a self-similar SGS model is  able to extend the anomalous scaling to almost the entire  range of resolved scales.  This  protocol reduces the computational cost by a  factor one thousand compared to a fully resolved DNS, with the same inertial range extension.  
The agreement between the scaling observed with the SGSM and that measured by DNS and experiments supports the universality of 
the inertial range dynamics with respect to the energy absorbing mechanism at small scales. {Thanks to the unprecedented accuracy in 
the determination of the scaling properties we are able to find some small discrepancy between the numerical data and the predictions by some of the most popular phenomenological models \cite{she1994universal,eling2015anomalous,yakhot2001mean} for high order moments. It remains an open key question to check if our closure remains accurate also at higher resolution. If this is indeed the case, we have a chance to make a discontinuous improvement in the assessment of scaling properties in homogeneous and isotropic turbulence.}  
Our model outperforms other common closures such as the Smagorinsky model or hyper-viscous DNS. Fully time-reversible models might of theoretical interest for the application of the chaotic hypothesis \cite{gallavotti1996extension}.
{Beside the self-similar properties, another advantage of our SGS closure is that the phase dynamics is left untouched.} Because of its generality, the closure can be applied to a broad set of other flow configurations such as rotating, stratified, or magnetohydrodynamic turbulence, including stiff problems as the kinematic dynamo in the limit of small Prandtl numbers \cite{dynamos}. {Similarly, one might imagine applications to wall bounded flows where small-scale anisotropy is negligible \cite{biferale2005anisotropy}, by imposing scaling laws on the spectral degrees of freedom in planes parallel to the wall (homogeneous directions), with properties dependent on the distance from the wall, eventually. }\\
\noindent We acknowledge useful discussions with  R. Benzi, M. Bustamante, M. Linkmann, C. Meneveau, Y. Oz, M. Sbragaglia, K.R. Sreenivasan, P.K. Yeung, M. Wilczek and funding from the European Union’s Programme (FP7/2007-2013) grant  No.339032. L.B. acknowledges the hospitality of the Center for Environmental and Applied Fluid Mechanics at Johns Hopkins University where this work was started. Part of the simulations have been done at BSC-CNS (PRACE Grant No. 188513). 

\section{Appendix: supplemental material `Self-similar subgrid-scale models for inertial range turbulence'}

\subsection{High order SF and error calculation}
In this supplemental material (SM) we further extend the quantitative analysis
of local scaling properties for high order generalized Flatness whose scaling properties
are summarized in  table II of the main body of the paper and in table I of this SM.
To quantify the  accuracy of the fit for $\bar \Delta_n$  we have measured the errors
in two different ways. The first, consists in the maximum difference between the exponents averaged on the whole inertial range, $[r_{min} = 0.03L: r_{max}=0.9L]$, with the values measured from the average on the first and the second half of the inertial range. The intermediate scale used in the evaluation of this error is $r_{int} = 0.18L$ obtained as $\log(r_{int}) = (\log(r_{min})-\log(r_{max}))/2$. The second  is the root-mean-square error, hence it is the sum of squared deviations between the fitted value $\bar{\Delta}_4$ and the data points weighted by the number $N$ of measurements used in the fit, namely, $\sqrt{\frac{1}{N} \sum_{i=0}^{N} (\bar{\Delta}_4 +2 - \xi_4(r_i)/\xi_2(r_i))^2 }$.\\}
{In Table I of the  paper we also show the prediction from three popular phenomenological models, the one proposed by She-Leveque (SL), Yakhot (Ya) and Eling-Oz (EO), which are given by the following expressions: 
\begin{equation}
\begin{cases}
\zeta_n = n/9 + 2(1-(2/3)^{n/3});\qquad \text{(SL),}\\
\zeta_n = [(\gamma-1)+\sqrt{(\gamma-1)^2+4\gamma n/3}]/(2\gamma);\qquad \text{(EO)},\\
\zeta_n = n (1+3\beta)/(3+3\beta n));\qquad \text{(Ya).}
\end{cases}
\end{equation}
All three expressions gives $\zeta_3=1$ independently of their free parameters. Both EO and Ya formula have one free parameter $\gamma$ and $\beta$, respectively. For each model the free parameter has been chosen such as  to match the mean value $\bar \Delta_4+2 = 1.843$ of the SGSM data. In order to evaluate the sensitivity of the model prediction  we give in Table II of the main paper the maximal variations obtained by fixing the free parameters to match either the maximum value measured within error bars by the SGSM model, $1.843 + 0.015$ or the minimum value $1.843-0.015$.} \\
{In Fig.~\ref{fig:5} we show the equivalent of Fig. 3 of the main paper but for $n=8$ and $n=10$. In the main body of the figure we report the log-lin plot of the local scaling exponents, $\Delta_n(r)+n/2= \xi_n(r)/\xi_2(r)$,  defining the scale-by-scale property of  $F_n(r)$ and in the inset the log-log of the Flatness. Notice that the SGSM closure allow us to achieve  good enough statistics  up to  $n=10$ (right panel). 
\begin{figure}[h]
  \includegraphics[scale=0.5]{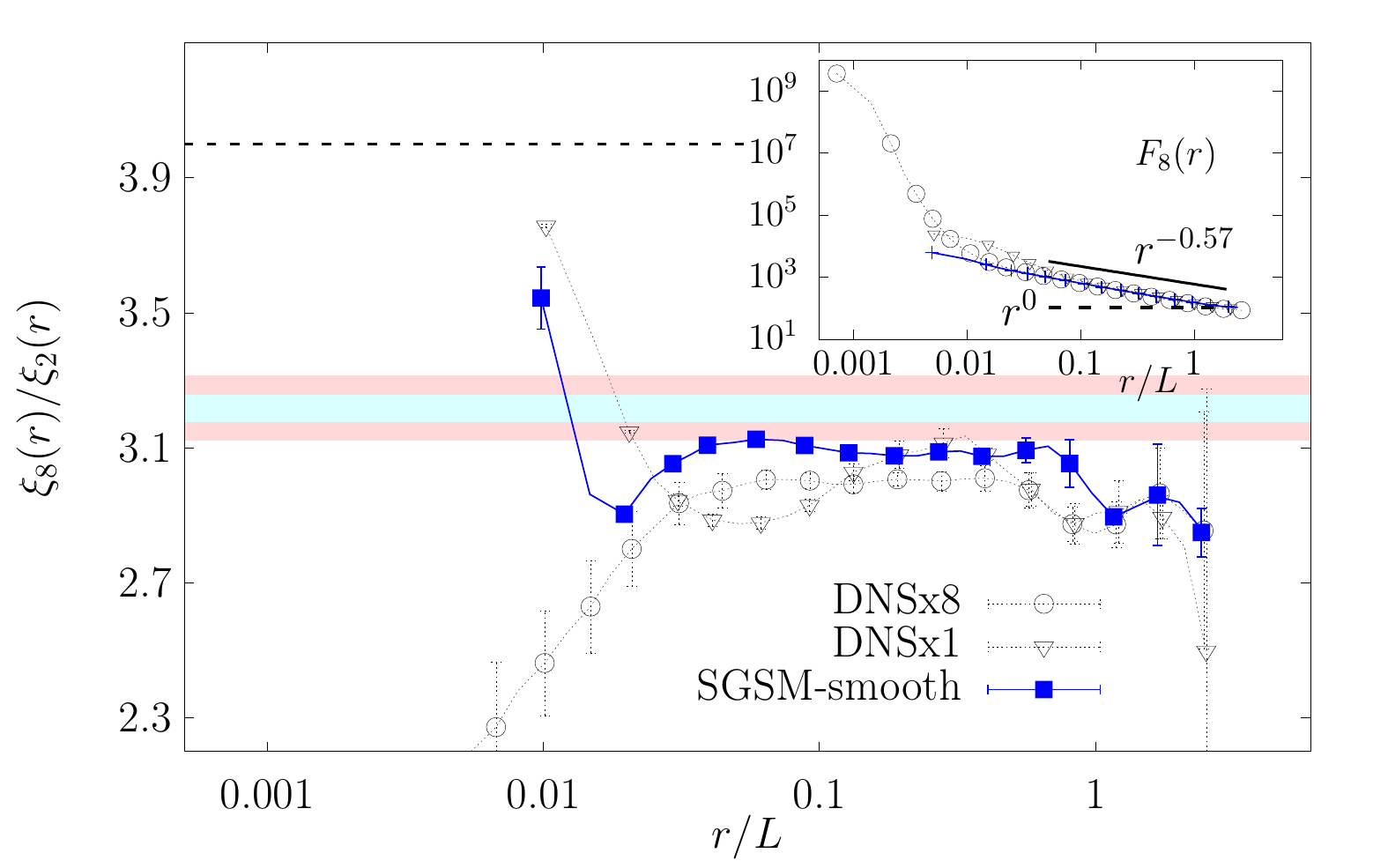}
  \includegraphics[scale=0.5]{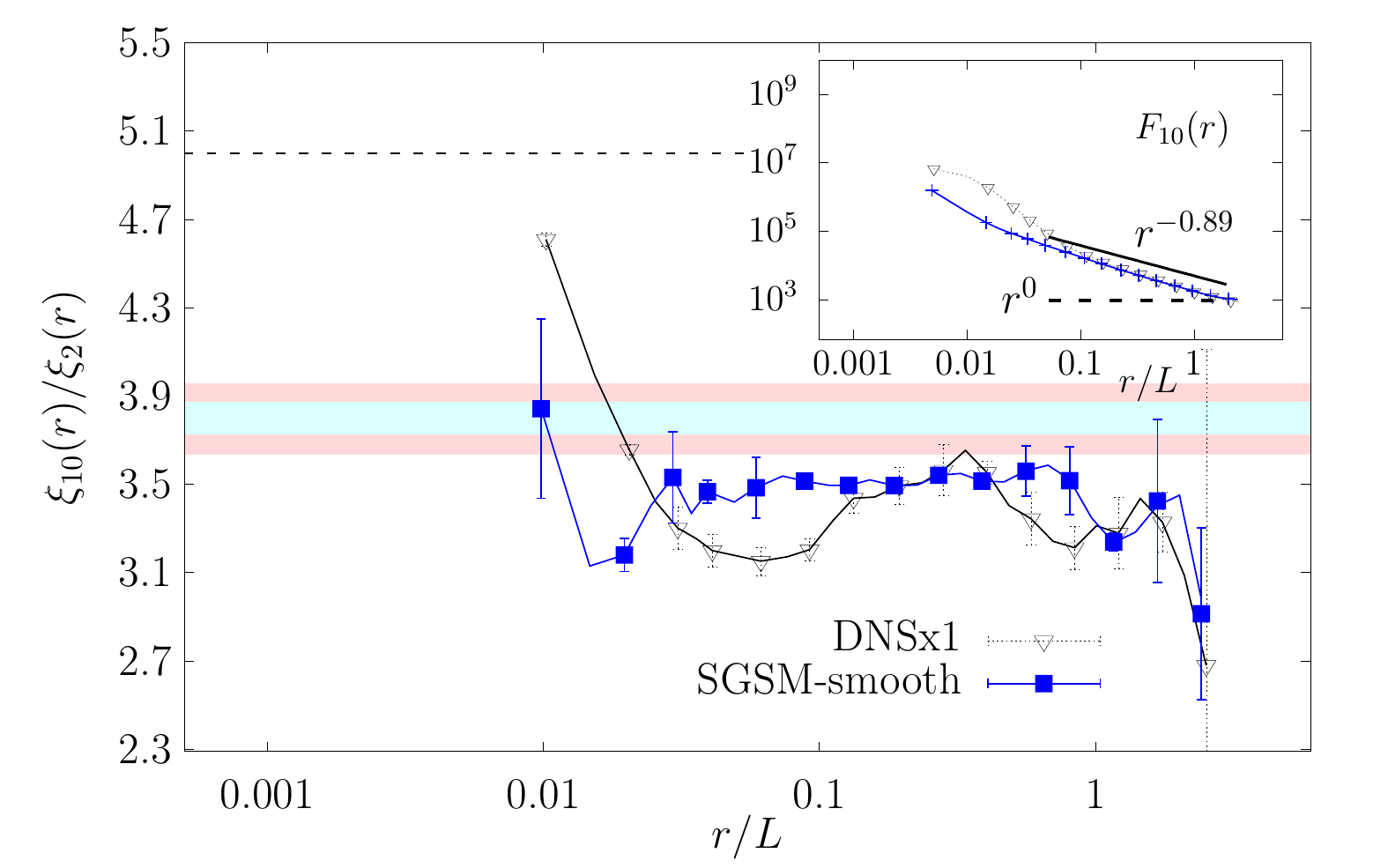}
\vspace{-0.3cm}
\caption{Left: log-lin plot of the local exponent, $\xi_8(r)/\xi_2(r)$, for both SGSMs and DNSs. The dashed horizontal lines represent the K41 predictions, 4. {The horizontal cyan bars indicate the values contained between the smallest and the highest prediction from the three models SL-Ya-EO in Table II of the main paper. The red horizontal bars show the same quantities including the maximal and minimal oscillations in the models predictions.} The value obtained from a fit of the SGSM-smooth model is $3.092 \pm 0.018 \pm 0.012$, where the two errors are estimated in two different ways (see table I below).
  Inset: log-log plot of Flatness vs $r$ with same symbols of the main
  panel.  The She-Leveque scaling, $\zeta_8-4\zeta_2=-0.57$, and the K41 prediction are given by the solid and the dashed lines, respectively.
  In all figures, errors are evaluated from the scatter of 40 configurations taken equispaced in time for a total length of 8-10 eddy turn over times. 
  Right: the same as left but for $n=10$.  Dashed horizontal line represents the K41 prediction, $5$. 
  The value obtained from a fit of the SGSM-smooth model is $3.504 \pm 0.050 \pm 0.031$. Here the DNSx8 data are missing because of lack of statistical accuracy. 
  }
\label{fig:5}
\end{figure}

\begin{table}
    \begin{tabular}{|c|ccc|c}
     & $N$ & $k_c$ & $k_{max}$  & $\bar{\Delta}_4+2$\\
    \hline 
    SGSM-sharp         & $1024$ & 340 & 512 & 1.837 (0.013)(0.013)  \\
    SGSM-sharp-dealias & $1024$ & 230 & 340 & 1.838 (0.006)(0.010) \\
    SGSM-smooth        & $1024$ & 340 & 512 & 1.843 (0.006)(0.009)  \\
    \end{tabular}
    \caption{ Parameters of simulations. 
    $N$: number of collocation points in each spatial direction; 
    $k_c$: smallest wavenumber where the SGS closure acts. 
    $k_{max}$: maximum wavenumber evolved by the dynamics.
    $\bar{\Delta}_4+2$: fit of the exponents $\xi_4(r)/\xi_2(r)$ shown in Fig. \ref{fig:7} in the range $r \in [0.03:0.9]\,L$}
    \label{tbl:table1}
\end{table}

\subsection{dealiasing effect}
As discussed in the main body of the paper, the SGSM-closure is not unique. One can play with (i)  the ratio $k_c/k_{max}$, (ii) the absolute value of $k_{max}$ and the distribution of wavenumbers where the spectrum is fixed inside the window $[k_c,k_{max}]$.  In Fig.~\ref{fig:7} we show the scale-by-scale ratio $\Delta_n(r)+n/2= \xi_n(r)/\xi_2(r)$ for $n=4$, for three different SGSM protocols (see table I in this SM):  without dealiasing ($k_{max}=N/2$, $k_c/k_{max} = 2/3$) for both SGSM-smooth and SGSM-sharp and with dealiasing ($k_{max}=N/3$, $k_c/k_{max} = 2/3$) for SGSM-sharp.  As we can see from Fig.~\ref{fig:7}, all simulations are in a very good agreement, suggesting that the aliasing errors are negligible in the estimation of the inertial-range scaling exponents with the self-similar SGSM closures explored here.
\begin{figure}[h]
\includegraphics[scale=0.55]{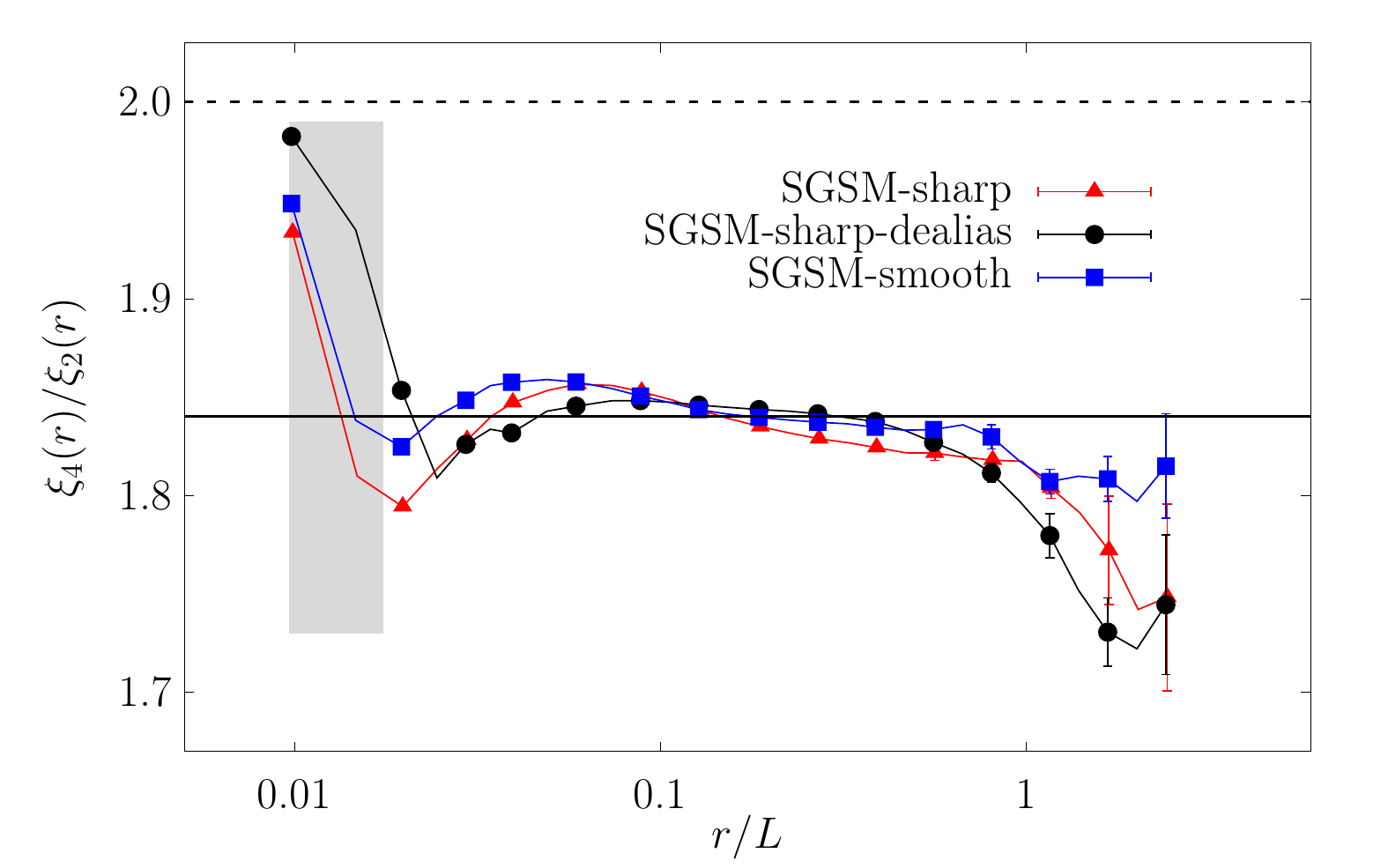}
\vspace{-0.3cm}
\caption{Comparison of $\xi_4(r)/\xi_2(r)$ for the simulations (i) SGSM-smooth, (ii) SGSM-sharp and (iii) SGSM-sharp correctly de-aliased. All simulations are performed with $1024^3$ collocation points. In the latter simulation (iii), the aliasing has been removed following the 2/3 rule by fixing $k_{max}=340$ and keeping the self-similar SGSM-sharp closure acting in the range $k_c/k_{max}=2/3$.  The SL anomalous scaling and the K41 prediction are given by the solid and the dashed lines, respectively.}
\label{fig:7}
\end{figure}
}

\bibliographystyle{apsrev4-1} 

\newpage

\end{document}